\documentclass[runningheads]{svmult}

\usepackage{makeidx}   
\usepackage{graphicx}  
\usepackage{subeqnar}  
\usepackage{multicol}  
\usepackage{physprbb}  
\makeindex             

\begin{document}
\title*{The disruption time of clusters in selected regions of four galaxies}
\toctitle{The disruption time of clusters in selected regions of four galaxies}
%
%
\titlerunning{The disruption time of clusters in four galaxies}
%
\author{Henny J.G.L.M. Lamers\inst{1,2}
\and Stratos G. Boutloukos\inst{1,3,4}}
\authorrunning{Henny J.G.L.M. Lamers \& Stratos G. Boutloukos}
%
%
\institute{
Astronomical Institute, Utrecht University, Princetonplein 5,
NL-3584CA, Utrecht, The Netherlands
\and
SRON Laboratory for Space Research
\and
Aristotle University, Thessaloniki, Greece
\and
University of Tuebingen, Institute of
 Theoretical Astrophysics and Computational Physics, Auf der
 Morgenstelle 10, D-72076, BRD
}

\maketitle              

\begin{abstract}
If the cluster formation rate is constant
and the disruption time of clusters depends on their initial mass as 
$t_{\rm dis} = t_4 \times (M_{\rm cl}/ 10^4 M_{\odot})^{\gamma}$,
the values of $t_4$ and $\gamma$ can be derived in a very simple way from the 
age and mass histograms of large homogeneous samples of clusters
with reliable age and/or mass determinations. 
We demonstrate the method and derive the values 
of $t_4$ and $\gamma$ from cluster samples in selected fields of four
galaxies: M51, M33, SMC and the solar neighbourhood.
The values of $\gamma$ are the same within their uncertainties 
in the four galaxies 
with $<\gamma>= 0.62 \pm 0.06$. However, the
disruption time $t_4$ of clusters of $10^4~M_{\odot}$ 
is very different in the different locations. It is shortest in the
inner region of M51 ($t_4 = 4 \times 10^7$ yrs) and highest in the SMC
($t_4=4\times 10^9$ yrs).
\end{abstract}

\section{Introduction}

Studies of the cluster formation history in different galaxies, 
based on {\it observed} cluster samples, rarely 
take into account the fact that clusters may have been disrupted.
We have started a series of studies to disentagle the effects of
cluster formation and cluster disruption. Here we present the method
for the determination of the disruption time in galaxies with a nearly
constant cluster formation rate,
with an application to selected regions in four galaxies.

Oort [12] noticed the lack of old open clusters in the solar
neighbourhood and derived statistically that Galactic clusters disrupt
on a timescale of $5 \times 10^8$ years. Several authors have found
that the mean age of clusters in the LMC and SMC is larger than that
of clusters in the solar neighbourhood and cloncluded that the
disruption time must be longer in the MCs (e.g. [9]).
Actually, the disruption time of clusters is expected to depend on
their mass and on the local conditions in their host galaxy.
Modern observations of large homogeneous cluster samples, e.g. with
$HST$, enable the determination of the cluster disruption time 
in different galaxies.

The disruption time of clusters is expected to scale with the
relaxation time 

\begin{equation}
t_{\rm dis} \sim t_{\rm rxt} \sim R_{\rm cl}^{3/2} M_{\rm cl}^{1/2}
\label{eq:trxt}
\end{equation}
where $R_{\rm cl}$ is the cluster radius and $M_{\rm cl}$ is the
cluster mass (see e.g. [13],[14]).
If all clusters in a limited region of a galaxy had the same radius,
then Eq. (1) would imply that $t_{\rm dis} \sim M_{\rm cl}^{1/2}$.
If the cluster radius depends on its mass as $R_{\rm cl} = C_{\rm gal} \times
M_{\rm cl}^{\delta}$, then we expect $t_{\rm dis} \sim C_{\rm gal}^{3/2} 
M_{\rm cl}^{\gamma}$ with $\gamma=(1+3\delta)/2$,
 where $C_{\rm gal}$ is a constant that
depends on the ambient pressure in the region of the galaxy where the 
clusters are located. (Obviously, this expected relation does not hold 
for clusters in highly elliptical orbits.)

In this paper we derive the empirical relation between the
disruption time and the mass of clusters in different locations
of galaxies, based on the mass and/or age distributions of 
magnitude limited cluster samples.

\section{The predicted mass and age distributions of 
surviving clusters above a given magnitude limit.}

Suppose that:

\begin{enumerate}
\item{} the {\it disruption time of clusters} depends on their initial mass $M_{\rm cl}$ as

\begin{equation}
t_{\rm dis}~=~t_4 \times (M_{\rm cl}/10^4~M_{\odot})^{\gamma}
\end{equation}
where $t_4$ is the disruption time of a cluster with an initial mass of
$10^4~ M_{\odot}$, 
\item{} the {\it formation rate of clusters} is constant,
\item{} the {\it initial mass function of clusters} can be written as 
$N(M_{\rm cl}) \sim M_{\rm
  cl}^{-\alpha}$, with $\alpha \simeq 2$ [16],
\item{}{\it clusters fade below the detection limit}, 
in the wavelength band that determines the 
magnitude limit of the cluster sample,
due to the evolution of their stars
as $F_{\lambda} \sim t^{-\zeta}$. The value of $\zeta$ can be derived
from cluster evolution models,
e.g. the Starburst99 models of Leitherer et al. [11].
\end{enumerate}

Then it is easy to show that both the age and mass distributions 
of surviving clusters above a certain magnitude limit 
will consist of double power laws (shown in Fig. 1) 
of the type:\\

\begin{itemize}

\item{} $ d N_{\rm cl} /d M_{\rm cl}  \sim  M_{\rm
  cl}^{(1/\gamma)-\alpha}$ ~~for low mass clusters due to  fading 
\item{} $d N_{\rm cl} / d M_{\rm cl} \sim  
 M_{\rm cl}^{\gamma-\alpha} $ ~~~~~~for high mass clusters due to disruption 
\item{} $d N_{\rm cl} / d t  \sim 
 M_{\rm cl}^{\zeta(1-\alpha)}$ ~~~~~~~for young clusters due to fading
\item{} $ d N_{\rm cl} /d t  \sim 
 M_{\rm cl}^{(1-\alpha)/\gamma} $ ~~~~~~for old clusters due to disruption 
\end{itemize}
These relations are derived by Boutloukos \& Lamers [4] (hereafter BL02).

The vertical shift of the fading lines in Fig. 1 depends on the
detection limit of the observations and on the number of clusters
formed per unit time in the region of the host galaxy that is studied.
 The horizontal shift of the
disruption line depends on $t_4$. The slope of the disruption line
depends on the values of $\gamma$ and $\alpha$. 
So both $t_4$ and $\gamma$ can be
derived from the cluster statistics in a simple straightforward way.

These double power laws are strictly valid if the clusters disrupt 
instantaneously when they reach an age of $t_{\rm dis}$. However, it can be
shown that gradual cluster disruption results in
similar power law distributions (BL02). In that case the transition 
between the two powerlaws is not a sharp bend, but it is more gradual.
This is demonstrated in Fig. 2. that shows the predicted mass and age
distributions in case the cluster mass decreases gradually on a
timescale that depends on the mass as in Eq. (2). The full lines are
the predicted relations for gradual disruption, whereas 
the dashed lines are the relations in case of
instantaneous disruption. They show a similar double power law as in
Fig. 1,  but with a gradual transition.

\begin{figure}
\includegraphics[width=6.0cm]{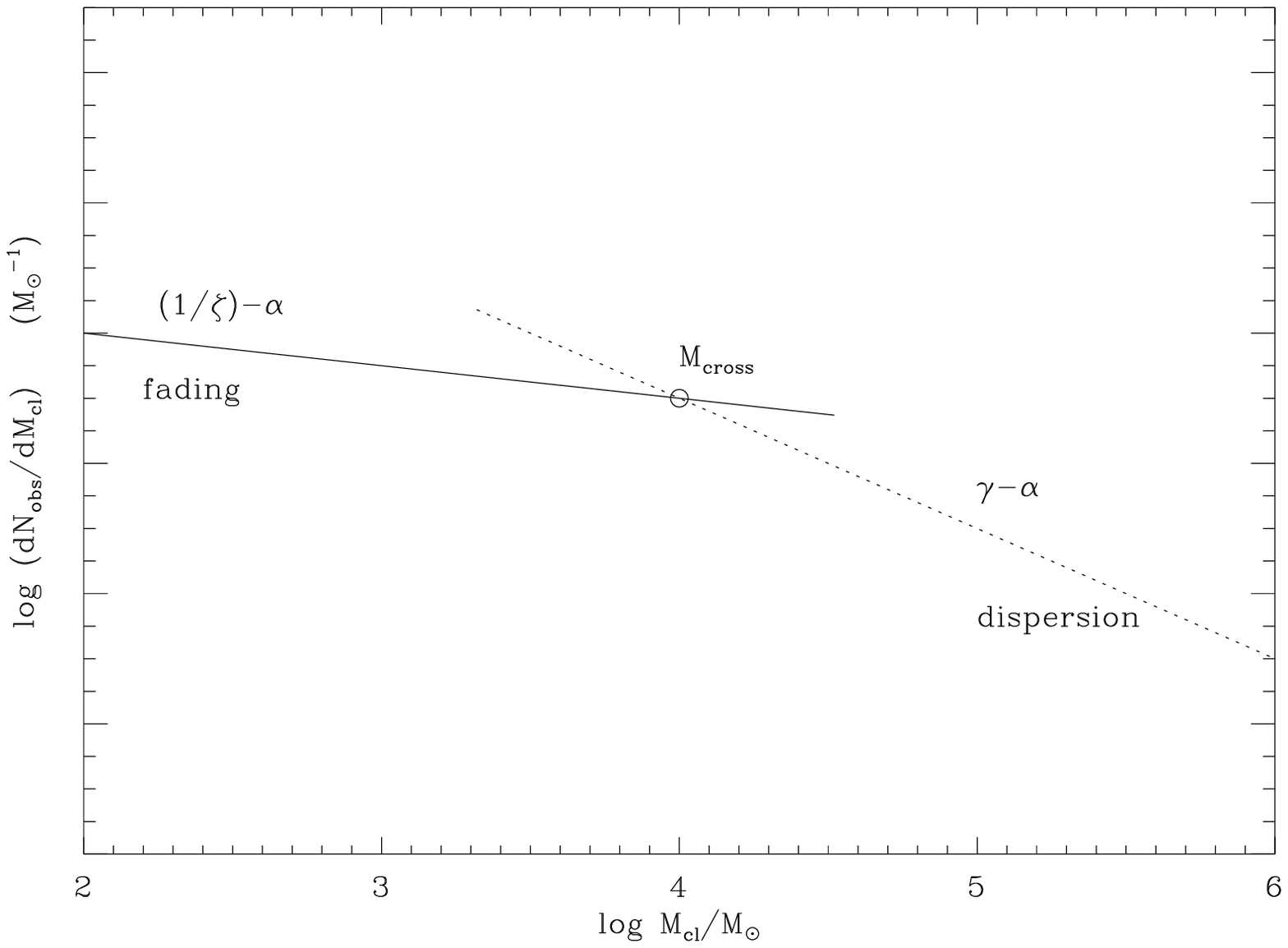}
\includegraphics[width=6.0cm]{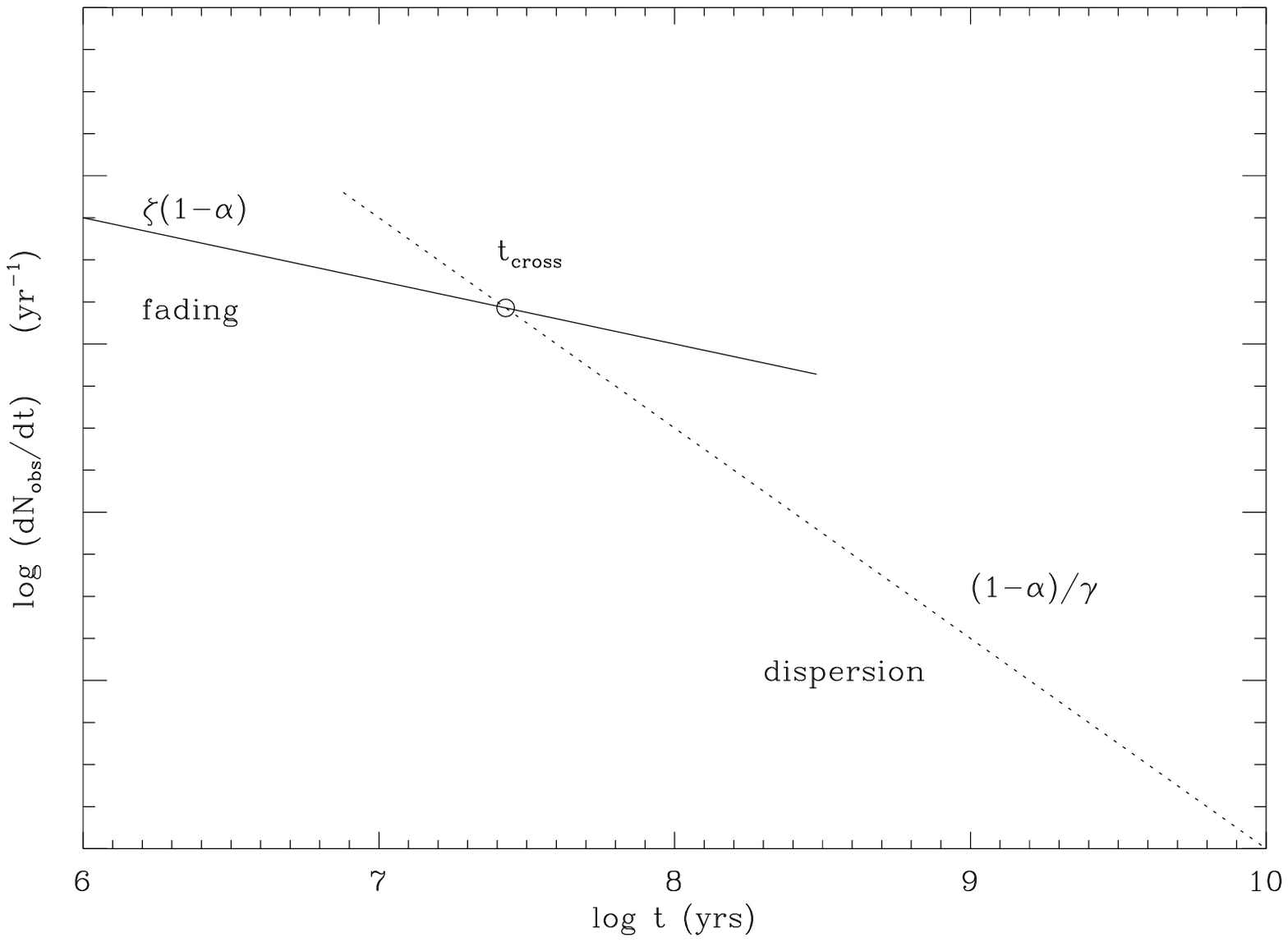}
\caption{Schematic pictures of the 
predicted mass distribution (left) and age distribution 
(right) 
of clusters due to fading (full lines) and
disruption (dotted lines). The $y$-axes have arbitrary units.
(figure from BL02)
} 
\label{fig:Npred}
\end{figure}
\begin{figure}
\includegraphics[width=6.0cm]{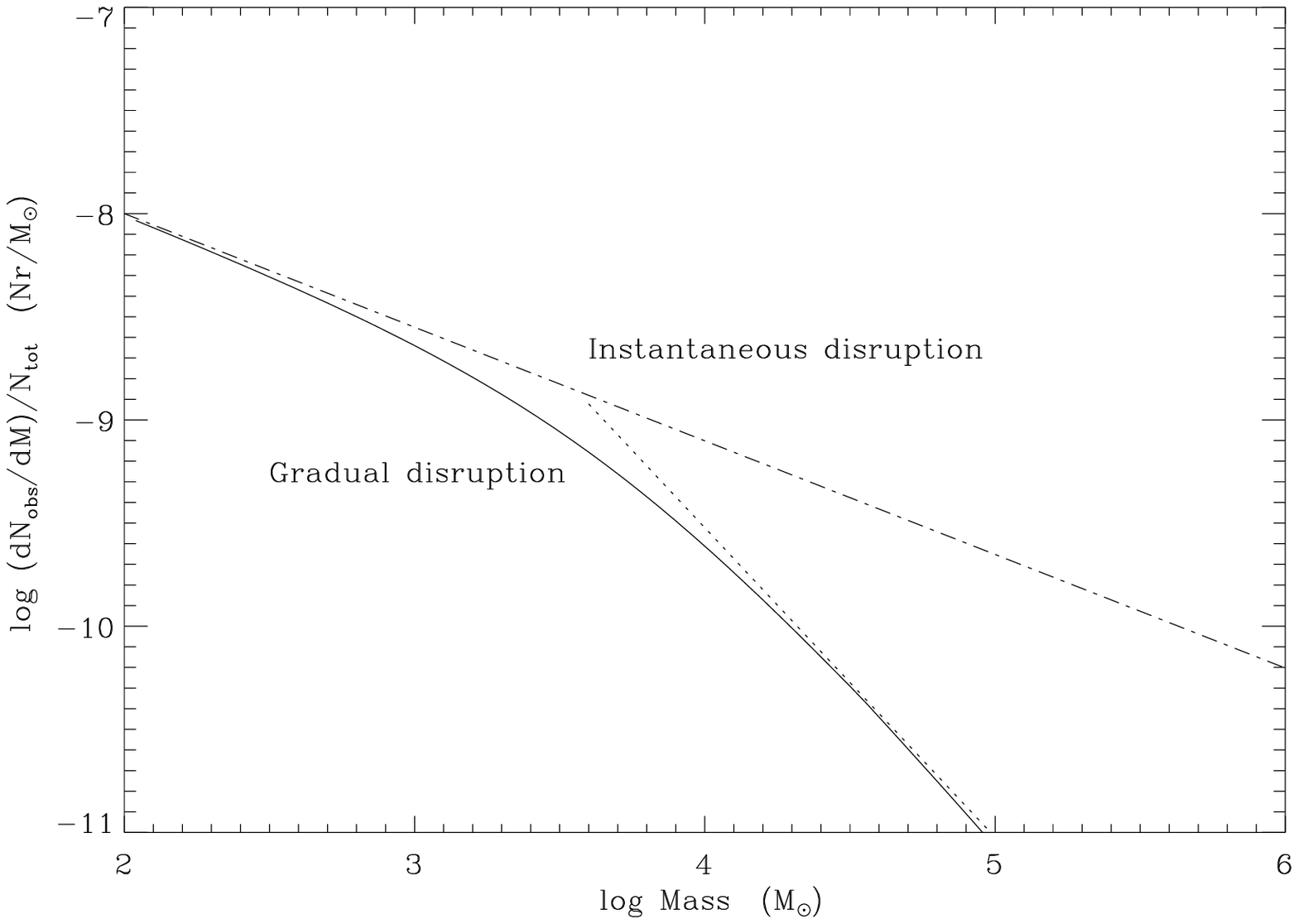}
\includegraphics[width=6.0cm]{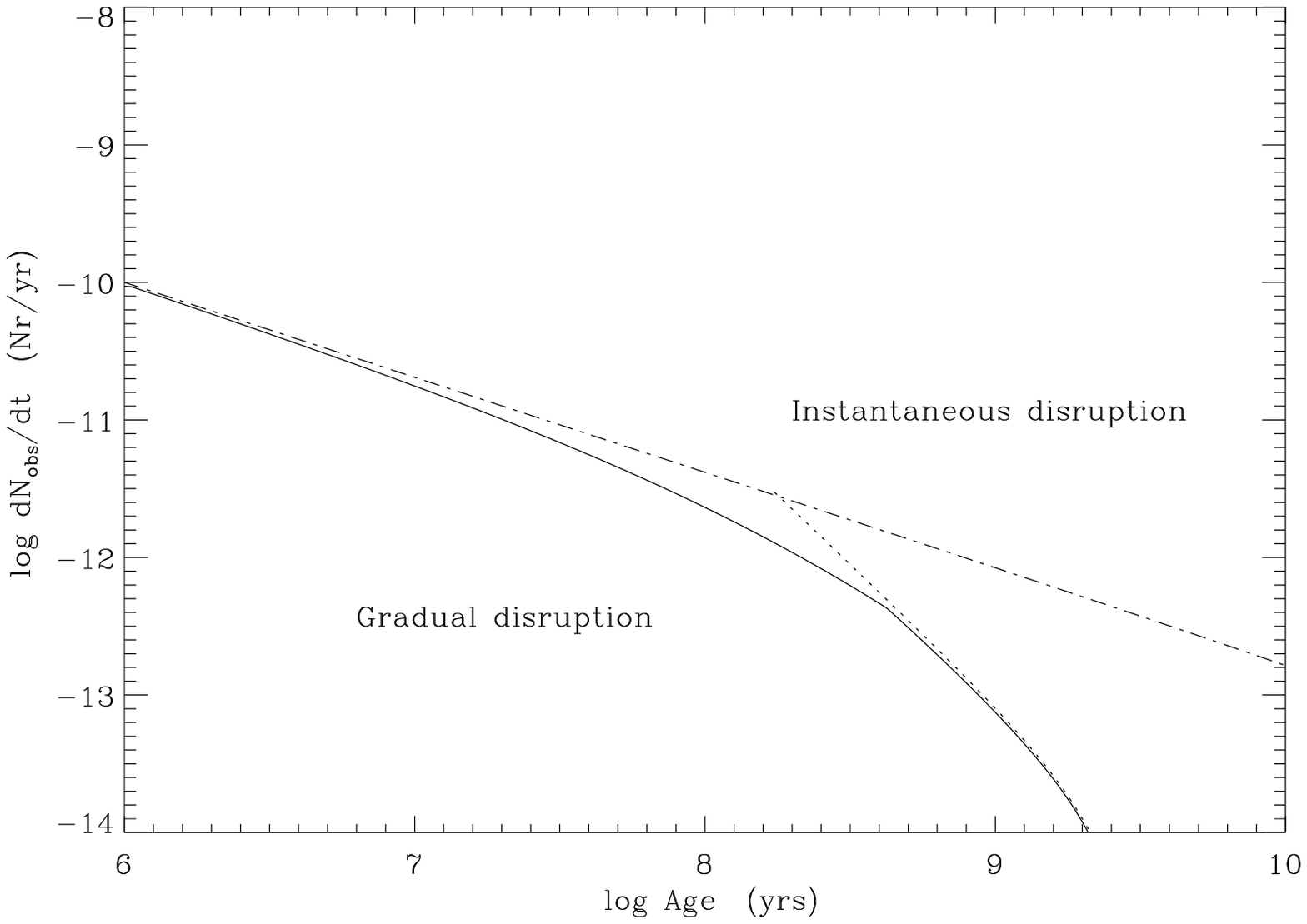}
\caption{Predictions for gradual decay of clusters with an age dependent
detection limit and with a disruption time that depends on the actual 
cluster mass as given by Eq. (2).
Left: the predicted mass distribution (full line).
Right: the predicted age distribution (full line).
The dash-dotted lines show the prediction if there was 
no disruption
but only evolutionary fading. The dotted lines show the prediction
for evolutionary fading and instantaneous disruption. The parameters
differ from those used in Fig. 1. (figure from BL02)}
\label{fig:appa}
\end{figure}

\section{Application to regions in four galaxies}

We have applied this method to samples of clusters observed with
$HST$ in the inner region of M51 by Bik et al. [3], 
in M33 by Chandar et al. [5],[6],[7] and
to the cluster samples of the SMC by Hodge [10]
and of the solar neighbourhood by Wielen [15]. The distributions
indeed show the power-law decrease at high mass and/or high age 
due to disruption. 
Figures 3a shows the mass distribution of the M51 
cluster sample and Fig. 3b shows the age distribution of the M33 
cluster sample. 
(BL02 also studied the age distribution of clusters in M51 and the
mass distribution of clusters in M33.)
The full lines are the predicted distributions for evolutionary fading
below the detection limit for these samples. (The wiggles at low age
or small mass are due to the appearance and disappearance of red
supergiants.) At high age or high mass 
the observed distributions
are powerlaws that clearly decrease much steeper than predicted for 
evolutionary fading only. This is due to disruption. 
Figures 4a and 4b show the age distributions of clusters in 
the SMC and the solar neighbourhood repectively. 
The age distribution of
the clusters in the SMC and the solar neighbourhood show a flat
part at young ages, because the clusters do not fade below the
detection limit in the samples used. Therefore, the distributions at
young age show the constant cluster formation rate. 
>From the observed powerlaw relations we derived the values of the disruption
parameters $t_4$ and $\gamma$.

The results are summarized in Table 1. 
The values of $\gamma$ are very similar (within the uncertainties)
for the cluster samples in the four galaxies, with a mean value of
$<\gamma>~=~ 0.62 \pm 0.06$.
The values of $t_4$, however, are very different for the different
galaxies: the inner region of M51 has the shortest 
cluster disruption time and the 
SMC has the longest disruption times. The difference amounts to
about a factor $10^2$.

\begin{table*}
\begin{center}
\caption[]{ The parameters of the disruption time:
$t_{\rm dis} = t_4 \times (M_{\rm cl}/ 10^4)^{\gamma}$}
\begin{tabular}{lcccccc}
\hline \\
Galaxy & Region    & Nr        & Age  & Mass  & $\log t_4$&
$\gamma$ \\
       &                  &           &range & range &           & \\
       & $r_{\rm gal}$(kpc)              & clusters  & log (yrs) & log $M_{\odot}$  & log (yrs) \\
\hline \\
 M51       & 0.8 - 3.1 &  512  & 6.0 -- ~9.7 & 3.0 -- 5.2 &
$7.64 \pm 0.22$  & $0.57 \pm 0.10$  \\
 M33       & 0.8 - 5.0 &  49  & 6.5 -- 10.0 & 3.6 -- 5.6 &
$8.12 \pm 0.30$  & $0.72 \pm 0.12$  \\
 MW & 7.5 - 9.5 &   72 & 7.2 -- 10.0 &  --- &
$9.0 \pm 0.3$   & $0.60 \pm 0.12$  \\
 SMC       & 0 - 2     &  314 & 7.6 -- 10.0 & --- &
$9.6 \pm 0.3$    & $0.61 \pm 0.08$  \\
 & & & & &      &   --------------- \\
 & & & & & Mean &   $0.62 \pm 0.06$  \\
\hline\\
\end{tabular}
\end{center}
\end{table*}

\begin{figure}
\includegraphics[width=6.0cm]{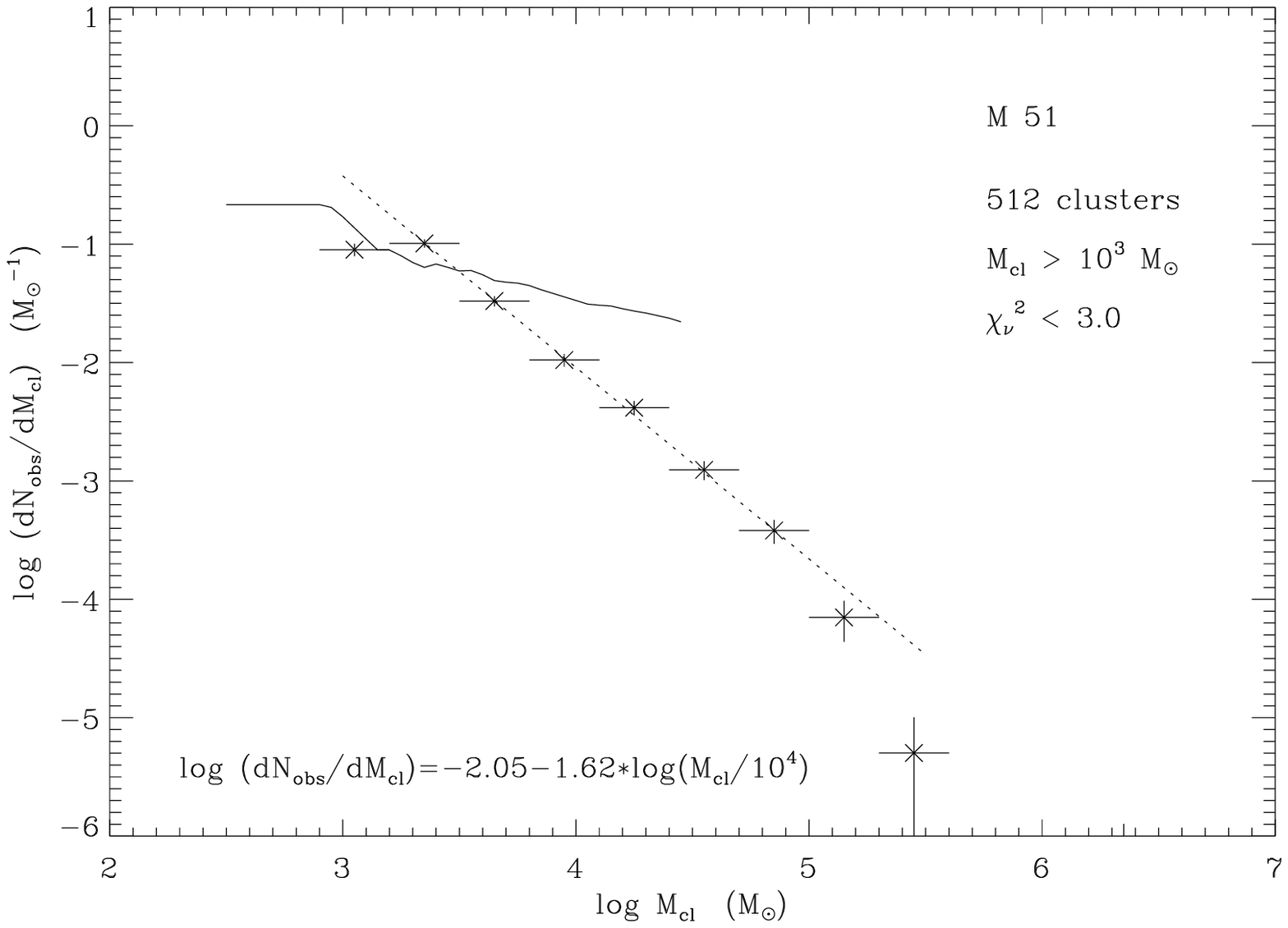}
\includegraphics[width=6.0cm]{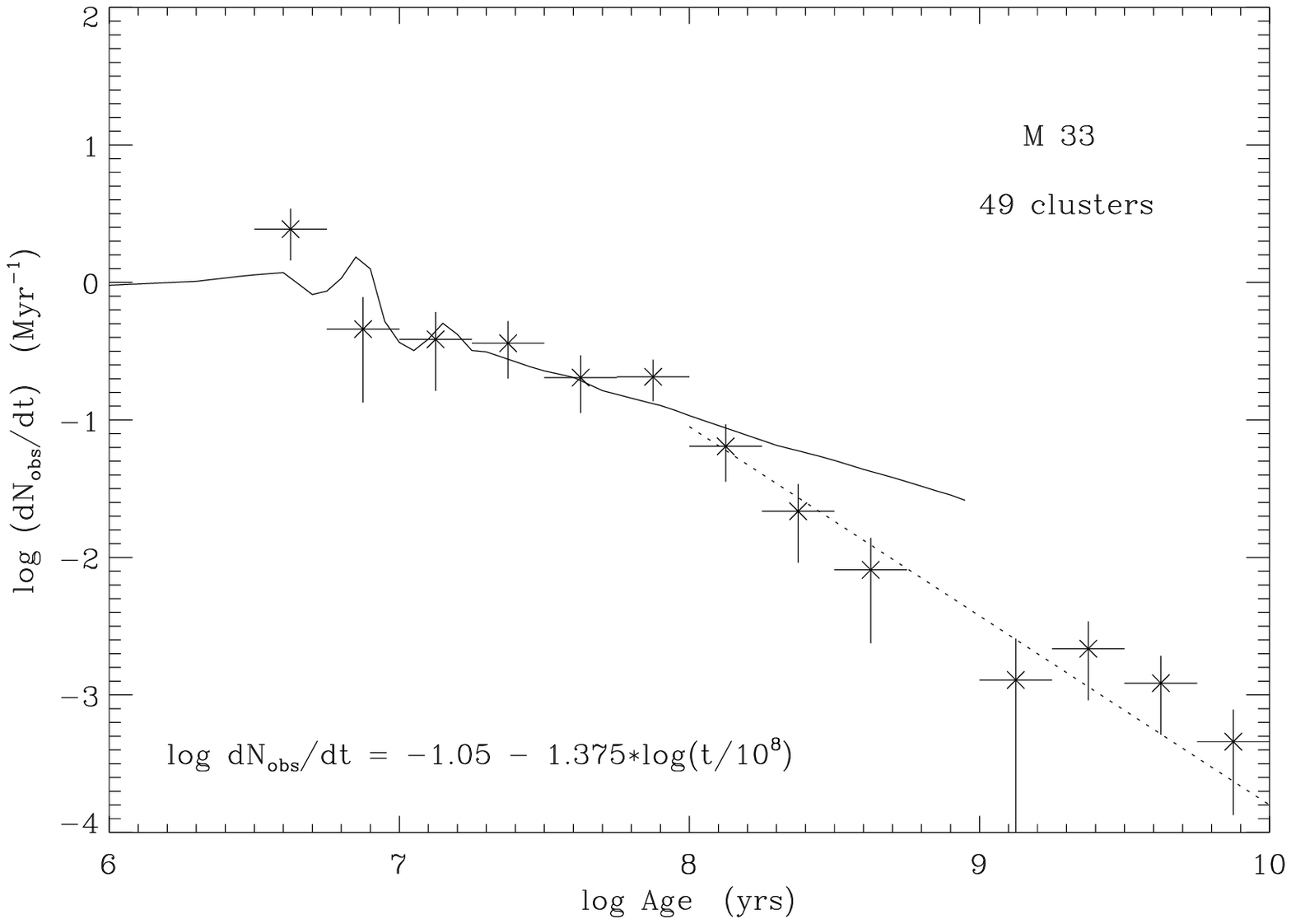}
\caption{The mass distribution of M51 clusters in the inner
0.8 - 3.1 kpc (left) and the age 
  distribution of M33 clusters at $0.8<r_{\rm Gal}<5.0$ kpc (right). 
The full lines are the predicted decrease
due to fading below the detection limit (from Starburst99 models). 
The dashed lines are powerlaw fits for disruption.
(figure from BL02)}
\label{fig:M51M33}
\end{figure}

\begin{figure}
\includegraphics[width=6.0cm]{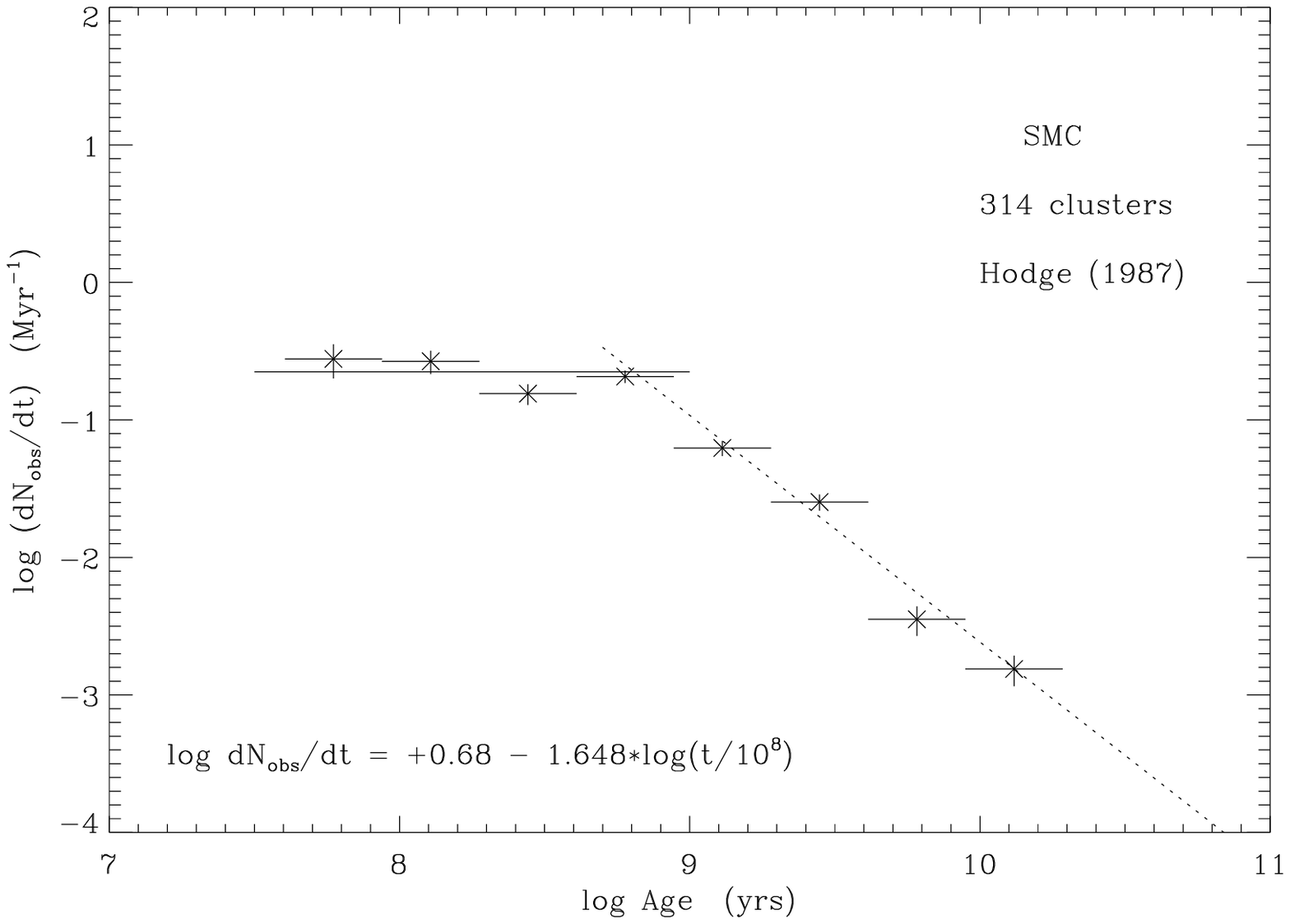}
\includegraphics[width=6.0cm]{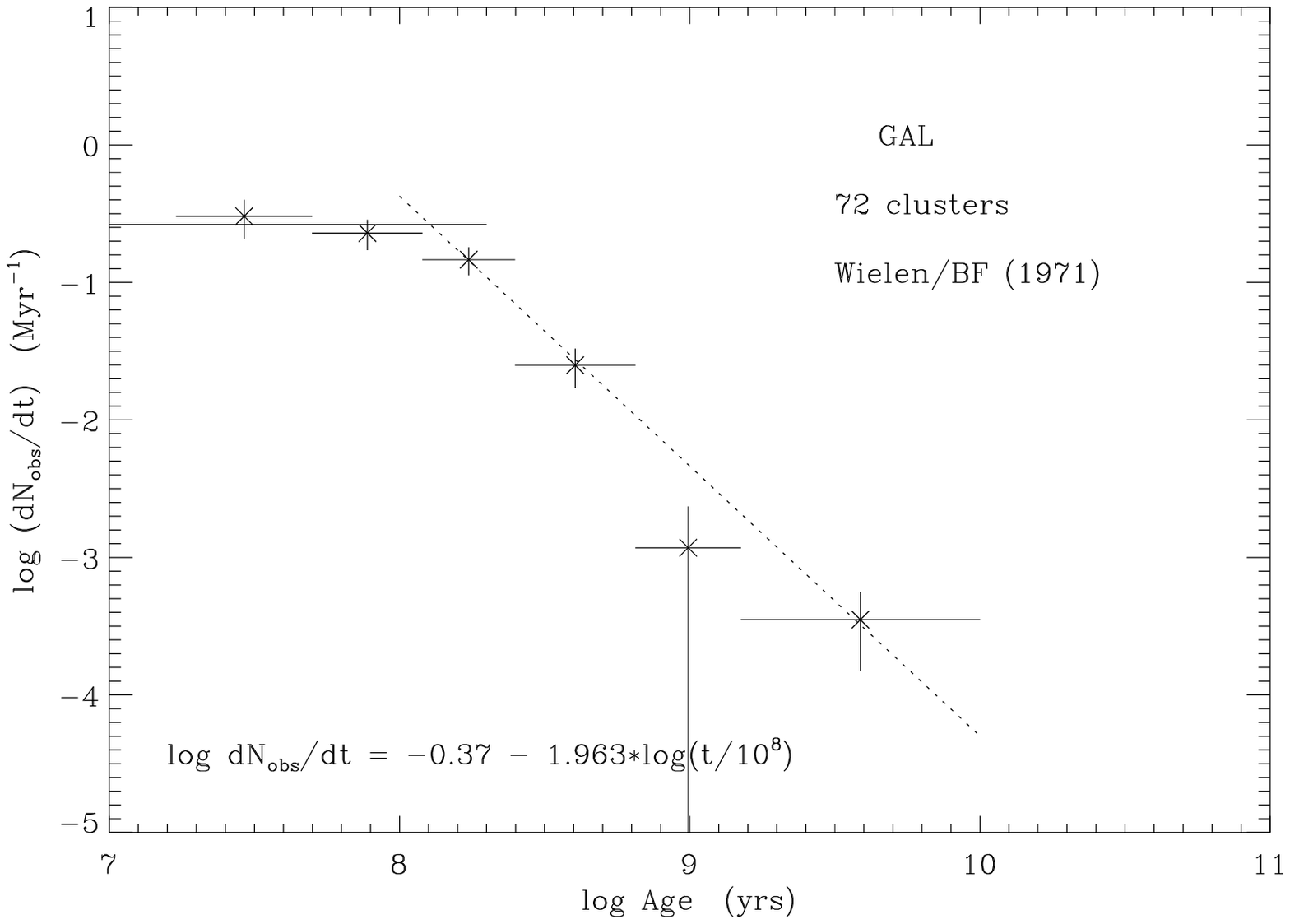}
\caption{The age distributions of clusters in the SMC (left)
and in the solar neighbourhood (right).
Evolutionary fading below the detection limit is
not important for these cluster samples, so the young
clusters show a constant cluster formation rate.
The dashed lines are powerlaw fits for disruption.
(figure from BL02)}
\label{fig:SMCGAL}
\end{figure}


\section{Future studies}

The full study has been described by BL02. Future planned or ongoing
studies include:
\begin{itemize}
\item{} study of the disruption time of clusters in different regions of M51
  (Bastian \& Lamers, [1])
\item{} study of the disruption time of clusters in the 
  starburst galaxy M82 (de Grijs et al. [8])
\item{} study of the disruption time of clusters in M81 
  as a function of mass and radius (Bastian et al., [2])
\item{} comparison of the derived disruption times with N-body simulations
  (Lamers \& Portegies Zwart, in preparation)
\item{} comparison of the disruption times of interacting and
  non-interacting galaxies (Bastian et al., in preparation)
\end{itemize}


\section*{Acknowledgements}

We thank Arjan Bik and Nate Bastian for their contribution to the 
analysis of the M51 $HST$-data. 
The stay of Stratos Boutloukos at the Astronomical Institute in
Utrecht was supported by a grant from
the Erasmus student project of the European Community.


%

\end{document}